\documentclass[
showpacs,
floatfix,
aps,
prl,
twocolumn,
superscriptaddress,
amssymb,
]{revtex4-1}

\usepackage{amssymb}
\usepackage{latexsym}
\usepackage[dvips]{graphicx}
\usepackage{amsmath}

\usepackage{graphicx}
\usepackage{dcolumn}
\usepackage{amsfonts}
\usepackage{bm}
\usepackage{epsfig}
\newcommand{\be}{\begin{equation}}
\newcommand{\ee}{\end{equation}}
\newcommand{\bea}{\begin{eqnarray}}
\newcommand{\eea}{\end{eqnarray}}

\def\pc{\mathcal{P}}

\newcommand{\wc}{\omega_{\rm c}}

\def\F{\mathcal{F}}

\def\eac{\epsilon}

\def\ttr{\tau}

\def\tsh{\tau_{\it sh}}
\def\tsm{\tau_{\it sm}}

\def\tin{\tau_{\it in}}

\def\oc{\omega_{\mbox{\scriptsize {c}}}}

\def\tq{\tau_{\mbox{\scriptsize {q}}}}
\def\tqdis{\tau_{\mbox{\scriptsize {q0}}}}

\def\ttr{\tau}

\def\tsh{\tau_{\mbox{\scriptsize {sh}}}}
\def\tsm{\tau_{\mbox{\scriptsize {sm}}}}

\def\tin{\tau_{\mbox{\scriptsize {in}}}}

\newcommand{\req}[1]{Eq.\,(\ref{#1})}

\newcommand{\rfig}[1]{Fig.\,\ref{#1}}

\newcommand{\rref}[1]{Ref.\,\onlinecite{#1}}

\usepackage[english,american]{babel}
\begin{document}

\title{Fine structure of high-power microwave-induced resistance oscillations}

\author{Q.~Shi}
\affiliation{School of Physics and Astronomy, University of Minnesota, Minneapolis, Minnesota 55455, USA}
\author{M.~A.~Zudov}
\affiliation{School of Physics and Astronomy, University of Minnesota, Minneapolis, Minnesota 55455, USA}
\author{I.~A.~Dmitriev}
\affiliation{Department of Physics, University of Regensburg, 93040 Regensburg, Germany}
\affiliation{Max Planck Institute for Solid State Research, 70569 Stuttgart, Germany}
\affiliation{Ioffe Physical Technical Institute, 194021 St. Petersburg, Russia}
\author{K.\,W. Baldwin}
\affiliation{Princeton University, Department of Electrical Engineering, Princeton, New Jersey 08544, USA}
\author{L.\,N. Pfeiffer}
\affiliation{Princeton University, Department of Electrical Engineering, Princeton, New Jersey 08544, USA}
\author{K.\,W. West}
\affiliation{Princeton University, Department of Electrical Engineering, Princeton, New Jersey 08544, USA}

\begin{abstract}
We report on observation of a fine structure of microwave-induced resistance oscillations in an ultra-clean two-dimensional electron gas.
This fine structure is manifested by multiple secondary sharp extrema, residing beside the primary ones, which emerge at high radiation power.
Theoretical considerations reveal that this fine structure originates from multiphoton-assisted scattering off short-range impurities.
Unique properties of the fine structure allow us to access all experimental parameters, including microwave power, and to separate different contributions to photoresistance.
Furthermore, we show that the fine structure offers a convenient means to quantitatively assess the correlation properties of the disorder potential in high-quality systems, allowing separation of short-range and long-range disorder contributions to the electron mobility.
\end{abstract}

\received{\today}
\maketitle

When a two-dimensional electron gas (2DEG) is exposed to microwave radiation and weak perpendicular magnetic field $B$, the magnetoresistance acquires prominent $1/B$-periodic oscillations \citep{zudov:2001a,ye:2001} which give rise to zero-resistance states \citep{mani:2002,zudov:2003,yang:2003,dorozhkin:2003,smet:2005,dorozhkin:2011,dorozhkin:2015,andreev:2003,auerbach:2005,alicea:2005,dmitriev:2013} when the resistivity at the minima approaches zero \citep{dmitriev:2012}.
These oscillations, termed microwave-induced resistance oscillations (MIRO), are usually \citep{note:altTheories,dmitriev:2004,beltukov:2016,volkov:2014,chepelianskii:2009,mikhailov:2011}
attributed to the effect of Landau quantization on the radiation-assisted scattering off impurities.
This affects the transport properties both directly (displacement mechanism \citep{ryzhii:1970,durst:2003,lei:2003,vavilov:2004,ryzhii:2004,khodas:2008}) and indirectly through the emergence of a nonequilibrium electron distribution function oscillating with energy (inelastic mechanism \citep{dmitriev:2003,dorozhkin:2003,dmitriev:2005}).
The resulting photoresistivity $\delta\rho_\omega$ oscillates with the ratio $\eac = \omega/\oc$ of the microwave frequency $\omega = 2\pi f$ to the cyclotron frequency $\oc=e B/m$ ($m$ is the electron effective mass) \citep{dmitriev:2009b},
\be
\frac{\delta \rho_\omega}{\rho_D} = - \lambda^2 \eta \pc\, 2\pi\eac \sin(2\pi\eac)\,,
\label{eq.miro}
\ee
where $\rho_D$ is the Drude resistivity, $\eta$ (specified below) is a dimensionless coefficient describing the combined strength of the displacement and inelastic contributions, $\pc$ is the dimensionless microwave power, and $\lambda=\exp({-\epsilon/2f \tq})$ is the Dingle factor.
In the expression for $\lambda$ ($\lambda \ll 1$), describing the amplitude of weak oscillations in the density of states due to Landau quantization, $\tq^{-1}=\tqdis^{-1}+\tin^{-1}(T)$ is the ``total'' quantum scattering rate including contributions from both the temperature ($T$) independent disorder and $T$-dependent electron-electron scattering ($\tqdis^{-1}$ and $\tin^{-1}$, respectively) to the Landau level broadening \citep{ryzhii:2004,dmitriev:2009b}.

Equation (\ref{eq.miro}), applicable in the regime of low power, $\pc \ll 1$, and overlapping Landau levels, $\lambda\ll 1$, well describes MIRO measured in many experiments \citep{dmitriev:2012}.
It has been demonstrated that under stronger microwave radiation, $\pc \gtrsim 1$, the oscillation amplitude undergoes a crossover to sublinear power dependence \cite{hatke:2011e}.
Concomitantly, the oscillation extrema move closer to the nodes at integer $\eac$ \cite{hatke:2011e}.
Such behavior is expected from theory \cite{vavilov:2004,dmitriev:2005,hatke:2011e} which attributes the nonlinear-in-$\pc$ corrections to multiphoton processes and/or to modification of thermalization rate.

In this Rapid Communication we report on a fine structure of MIRO which emerges under intense low-frequency microwave irradiation in an ultra-clean 2DEG.
This fine structure is manifested by additional (``secondary'') sharp extrema residing besides the primary ones.
Following theoretical framework of \rref{hatke:2011e}, we demonstrate that this fine structure originates from multiphoton-assisted scattering off short-range impurities, which are inherently present even in ultra-high mobility 2DEG.
The properties of fine structure enable us to determine the microwave power seen by our 2DEG and to single out different photoresistance contributions, what has not been possible before \citep{dorozhkin:2016}. The fine structure can also potentially be used as a tool to separate sharp and smooth disorder components of the electron mobility which is a subject of intense current interest but is difficult to achieve by conventional techniques \citep{umansky:1997,umansky:2013,manfra:2014,dassarma:2014}.

The data presented below were obtained from a $\sim 4$~$\times$~4 mm square sample cleaved from a heterostructure containing a symmetrically modulation-doped, 30-nm wide GaAs/AlGaAs quantum well.
A brief illumination with a light-emitting diode resulted in an electron density $n \approx 3.0 \times 10^{11}$~cm$^{-2}$ and a mobility $\mu \approx 3.1 \times 10^7$~cm$^2$/Vs (transport relaxation time $\ttr\approx 1.2$~ns).
The microwave radiation from the signal generator was delivered to the sample via a rectangular waveguide at various powers spanning two orders of magnitude.
The longitudinal resistance was measured using low-frequency lock-in amplification under continuous microwave irradiation at a pumped liquid helium bath temperature of $T \approx 1.35$~K.

\begin{figure}
\includegraphics{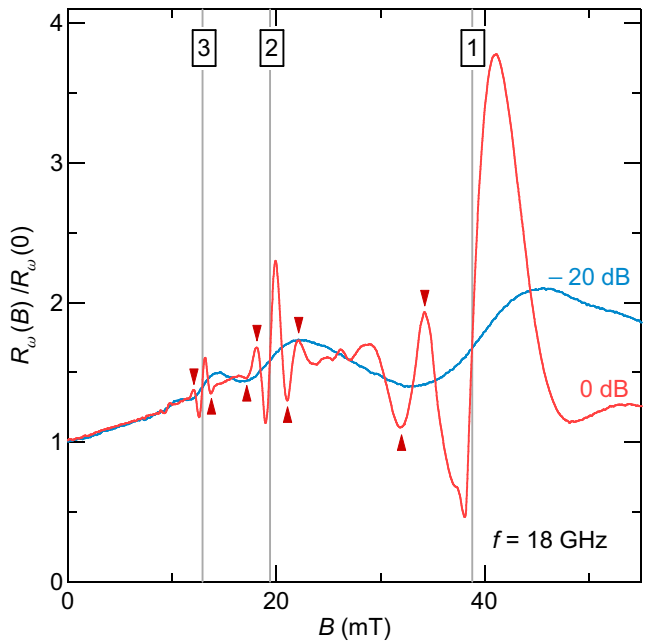}
%\vspace{-0.05 in}
\caption{(Color online)
Longitudinal magnetoresistance normalized to its zero-field value, $R_{\omega}(B)/R_{\omega}(0)$, measured at $f = 18$ GHz
for two different power levels corresponding to an attenuation $P_\text{dB}=0$~dB and $-20$~dB.
Vertical lines mark $\epsilon$ = 1, 2, 3.
Secondary extrema are marked by triangles.
}
\vspace{-0.15 in}
\label{fig1}
\end{figure}

In \rfig{fig1} we present the longitudinal magnetoresistance normalized to its zero-field value, $R_{\omega}(B)/R_{\omega}(0)$, recorded under microwave radiation of frequency $f = 18$~GHz for two different power levels corresponding to $0$~dB and $-20$~dB attenuation, as marked.
Vertical lines, marked by $\eac= 1, 2, 3$, correspond to cyclotron resonance harmonics characterized by vanishing photoresponse.
At low power ($P_\text{dB}=-20$~dB), we observe smooth MIRO exhibiting the conventional damped sine-like waveform, as prescribed by \req{eq.miro}.
In contrast, the data at the maximum power $P_{\max}$ ($P_\text{dB}=0$~dB) reveal a very rich and unusual waveform highlighted by multiple additional sharp features around each node at $\eac=N=1, 2, 3$.
The ``secondary extrema'', whose positions are marked
by triangles, lie roughly symmetrically around each node and have an amplitude comparable to the primary extrema closest to the nodes.
Though less pronounced, similar fine structure was observed at other frequencies \citep{note:101}.

\begin{figure}[t]
\includegraphics{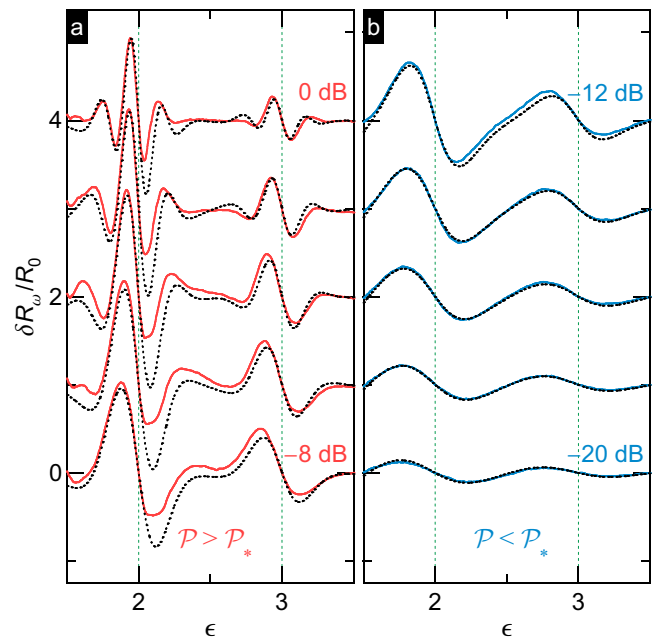}
%\vspace{-0.05 in}
\caption{(Color online)
Measured (solid lines) and calculated (dotted lines) photoresistance normalized by the zero-field resistance without radiation, $\delta R_{\omega}/R_0$, as a function of $\eac$ at different power attenuations: from 0 dB to $-8$ dB (a) and from $-12$ dB to $-20$ dB (b), in $-2$ dB steps.
The parameters of disorder used in calculation are: $\tau=1.2$~ns, $\tsh=1.7$~ns, and $\tsm=75$~ps.
}
\vspace{-0.15 in}
\label{fig2}
\end{figure}

In \rfig{fig2} we plot the measured photoresistance (solid lines) normalized to the zero-field resistance without radiation, $\delta R_{\omega}/R_0$, as a function of $\eac$ at different powers.
Here, we focus on structures around $\eac=2$ and 3 where Landau levels strongly overlap, $\lambda\ll 1$.
The high and low power regimes are illustrated in (a) and (b) respectively.
A gradual transition between the two distinct waveforms shown in \rfig{fig1} is apparent.
With decreasing power, secondary extrema move away from integer $\eac$. Their amplitude decreases until they disappear at $P_\text{dB}\sim -8$~dB (-8~dB corresponds to $\pc\simeq\pc_*$ specified below).
This behavior clearly distinguishes the observed fine structure from the fractional MIRO \cite{zudov:2004,zudov:2006a,dorozhkin:2007,pechenezhskii:2007,wiedmann:2009}
--- distinct features in the photoresponse around certain rational $\eac=1/2$, 1/3, 3/2, etc. characteristic for the regime of well separated Landau levels \cite{lei:2006b,dmitriev:2007b} and the crossover regime $\wc \tq\sim 1$ \cite{pechenezhskii:2007}. 
At still lower power [panel (b)] the positions of the primary extrema still move away from the nodes and finally saturate at $\eac=N\pm1/4$ consistent with \req{eq.miro}.

We now turn to a quantitative analysis of the results within the quantum kinetic framework developed in Refs.~\citep{vavilov:2004,dmitriev:2005,vavilov:2007,khodas:2008,dmitriev:2007,hatke:2011e}.
For arbitrary microwave power $\pc$ and for $\lambda\ll 1$, the general expression for the photoresponse reads \cite{hatke:2011e}
\begin{align} \label{c5a}
\frac{ \delta \rho_\omega }{ \rho_{D} } = \lambda^{2} (\F_\text{dis} +\F_\text{in}),
\end{align}
with the displacement and inelastic terms given by
\begin{align} \label{f}
\F_\text{dis}= \tau\partial_\eac [\eac\bar{\gamma}(\xi)]-1,\quad
\F_\text{in}=\frac{ 2 \eac\tau\bar{\gamma}(\xi)\partial_\eac \gamma(\xi) }{ \tin^{-1} + \gamma(0) - \gamma(\xi)},
\end{align}
where $\xi = 2 \sqrt \pc \sin \pi \eac $.
Aiming at a realistic description of elastic scattering in high-mobility 2DEGs with a minimal number of parameters, we use a two-component disorder model \cite{vavilov:2007}, with its sharp (smooth) component being weakly (strongly) correlated at the scale of the Fermi wave length. Within this model,
\begin{subequations}\label{gammainmodel=}
\bea
\label{DisModel17}
\gamma(\xi) &=& \frac{ J_0^2(\xi) }\tsh+ \frac 1 \tsm \frac 1 { (1 + \chi \xi^2)^{1/2}} \,,\\
\label{DisModel20}
\bar{\gamma}(\xi) &=& \frac{ 1 }{ \tsh }
\left[ J_0^2(\xi) - J_1^2(\xi) \right]
 +
\frac{ \chi }{\tsm } \frac{ 1 - \chi \xi^2 / 2}{(1 + \chi
\xi^2)^{5/2}},
\eea
\end{subequations}
where $\tsh^{-1}$ and $\tsm^{-1}$ ($\tsm^{-1}\gg\tsh^{-1}$) characterize the strength of sharp and smooth disorder, respectively, while $\chi^{1/2}\ll 1$ is the typical scattering angle off smooth disorder.
Here, the smooth component describes the potential of remote ionized donors.
It dominates in the quantum rate: $\tqdis^{-1}\equiv\gamma(0)=\tsm^{-1}+\tsh^{-1}\simeq\tsm^{-1}$.
In turn, the short-range  component accounts for residual impurities in close proximity to the 2DEG \cite{note:mixed} which can contribute significantly to the momentum relaxation rate, $\ttr^{-1}\equiv\bar{\gamma}(0)=\chi\tsm^{-1}+\tsh^{-1}$.
This model was successfully employed to describe related nonequilibrium effects in the presence of strong dc excitation \citep{yang:2002,zhang:2007a,zhang:2007c,vavilov:2007,khodas:2008,khodas:2010,wiedmann:2011c,chakraborty:2014}. With $\bar{\gamma}$ given by \req{DisModel20}, the displacement term $\F_\text{dis}=\F_\text{dis}^\text{(sh)}+\F_\text{dis}^\text{(sm)}$ entering \req{c5a} naturally divides into the ``sharp'' ($\propto\tsh^{-1}$) and ``smooth'' ($\propto\tsm^{-1}$) parts, while for $\F_\text{in}$ such separation is not possible.

\begin{figure}[t]
\includegraphics{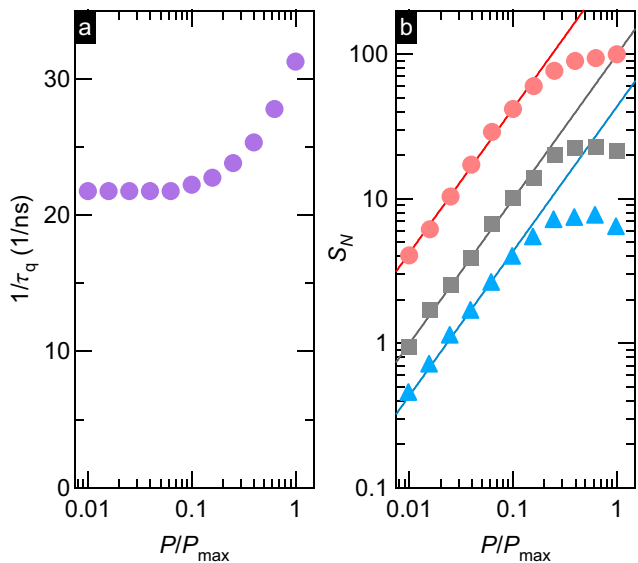}
%\vspace{-0.05 in}
\caption{(Color online)
(a) Extracted quantum scattering rate $1/\tq$ vs normalized power $P/P_{\max}$.
(b) Slope $S_N$, obtained from the data, as a function of $P/P_{\max}$, on a log-log scale, for $N=1$ (circles), 2 (squares) and 3 (triangles).
}
\vspace{-0.15 in}
\label{fig3}
\end{figure}

As argued below, the fine structure in Figs.~\ref{fig1} and \ref{fig2} originates from the sharp-disorder displacement contribution,
\be\label{FdisSh}
\F_\text{dis}^\text{(sh)}=-\dfrac{\ttr}{\tsh} \pi\eac\cot(\pi\eac)\xi J_1(\xi) \left[ J_2(\xi) - 3 J_0(\xi) \right],
\ee
The expression involving Bessel functions oscillates with $\xi$ \cite{note:omitted}.
Neglecting other contributions, for $\pc\gg \sin^{-2}(\pi\eac)$ ($\xi\gg 1$) one thus obtains
\be\label{FdisShCos}
\frac{ \delta \rho_\omega }{\rho_{D} } \simeq 4\lambda^2 \dfrac{\ttr}{\tsh} \cot(\pi\eac) \cos(4 \sqrt{\pc}\sin\pi\eac).
\ee
Equations (\ref{FdisSh}) and (\ref{FdisShCos}) predict that, with increasing $\pc$, more and more secondary extrema should show up in the photoresponse and, once present, each extremum should move towards the nearest node.
The sensitivity of the extrema positions to microwave power thus offers a unique opportunity to quantitatively access $\pc$ sensed by our 2DEG.
Indeed, the positions of principal secondary extrema are given by $\xi= 2 \sqrt \pc \sin \pi \eac\simeq \pm 2.9$.
Such scaling is observed in our data shown in \rfig{fig2} and the analysis using \req{FdisSh} for $P_\text{dB}>-8$~dB yields $\pc\simeq 15 P/P_\text{max}$.
We also notice that the secondary extrema are expected to emerge at $\pc>\pc_*=1.45^2\sim 2.1$ (as follows from $|\xi|\geq 2.9$ and $|\sin \pi \eac|\leq 1$).
In accord with this estimate, the fine structure in \rfig{fig2} becomes visible at $P_\text{dB}=-8$~dB which corresponds to $\pc\simeq 2.4$.

Having obtained $\pc$ as a function of power attenuation, we turn to examination of the data at low power, where Eqs.~(\ref{c5a})-(\ref{gammainmodel=}) reduce to the linear-in-$\pc$ \req{eq.miro} with
\be\label{lowP}
\eta\equiv\eta_\text{in}+\eta_\text{sh}+\eta_\text{sm}=\dfrac{2\tin}{\ttr}+\dfrac{3\ttr}{2\tsh}+\dfrac{6\chi^2\ttr}{\tsm}.
\ee
The inelastic lifetime can be estimated as $\tin(T) \simeq (\hbar E_F/T^2)\ln^{-1}(a_B v_F/2\Omega)$, where $E_F$ and $v_F$ denote the Fermi energy and velocity, $\Omega=\mathrm{max}\{T/\hbar,\omega_c^{3/2}\tau^{1/2}\}$, and $a_\text{B}\simeq 10$~nm is the effective Bohr radius \citep{dmitriev:2005}.
At the coolant temperature $T_0=1.35$~K, we obtain $\tin\simeq 120$~ps (at $\eac\sim 2$).
On the other hand, the Dingle analysis of the lowest power photoresistance ($P_{\rm dB} = - 20$ dB) yields the quantum lifetime $\tq \approx 46$~ps comparable to the estimate for $\tin$. Since $\tq^{-1}=\tqdis^{-1}+\tin^{-1}$, the contribution of electron-electron scattering to the Landau level broadening should be taken into account, wherefrom we obtain $\tsm\simeq\tqdis\simeq 75$~ps. We then model the lowest-$\cal P$ photoresistance trace using Eqs.~(\ref{eq.miro}) and (\ref{lowP}), with $\ttr/\tsh$ being the only adjustable parameter (note that $\chi$ can be found from $\ttr/\tsh+\chi\ttr/\tsm=1$ for a given $\ttr/\tsh$, $\ttr$, and $\tsm\simeq\tqdis$).
A dotted line, computed using $\ttr/\tsh = 0.7$, is virtually indistinguishable from the experimental trace \citep{note:errors}.
From this analysis we conclude that MIRO at low $\pc$ are dominated by the displacement term associated with the sharp disorder component, $\eta_\text{sh}\simeq 1.05$, $\eta_\text{in}\simeq 0.2$, and $\eta_\text{sm}\simeq 0.03$.

With a knowledge of $\ttr/\tsh$, we can now compute the photoresistance using Eqs.~(\ref{c5a})-(\ref{gammainmodel=}) for all other power levels.
The results, presented by dotted lines in \rfig{fig2}, demonstrate excellent agreement at all $P_{\rm dB}$, accurately capturing the fine structures observed at higher intensities. This supports the previous analysis and confirms that the detected fine structure of MIRO originates from the multiphoton-assisted scattering by short-range impurities as described by Eqs.~(\ref{FdisSh}) and (\ref{FdisShCos}).

Apart from the known microwave power, the only parameter which was allowed to vary while generating theoretical curves in \rfig{fig2} was the electron temperature, to account for electron heating by high-power microwaves. Since $\tin\sim\tq\ll\tau\sim\tsh$, we expect that heating affects primarily the quantum scattering rate entering the Dingle factor. The prefactor $\eta$, on the other hand, is affected only slightly, as it is dominated by $\eta_\text{sh}$.
Our analysis reveals that $\tq$ initially stays constant at $\tq(T_0)\approx 46$~ps but eventually decreases to $\tq(T_\text{max})\approx 32$~ps at $P=P_\text{max}$, see Fig.~\ref{fig3}(a).
Using $\tin(T_0) = 0.12$~ns and $\tin\propto T^{-2}$, we obtain $T_\mathrm{max}\simeq 1.7~\text{K}$ and $\tin(T_\mathrm{max})\simeq 0.5~\tin(T_0)$.

\begin{figure}[t]
\includegraphics{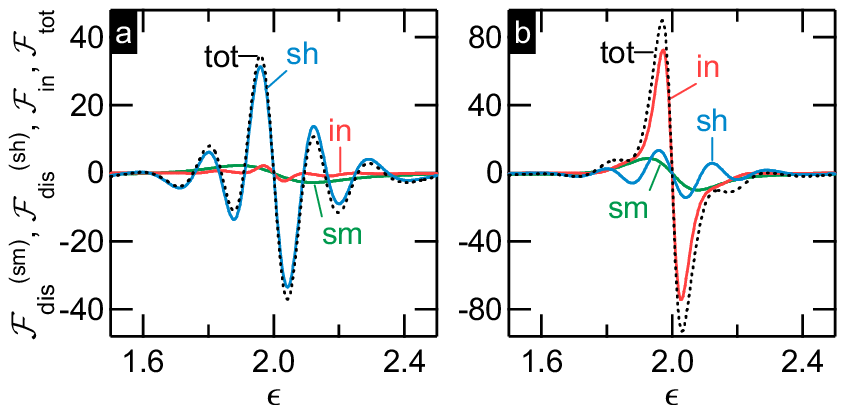}
%\vspace{-0.05 in}
\caption{(Color online) (a) Calculated $\F_\text{dis}^\text{(sh)}$ (marked as ``sh''), $\F_\text{dis}^\text{(sm)}$ (``sm''), $\F_\text{in}$ (``in''), and their sum $\F_\text{tot}$ (``tot'') vs $\eac$, for $\ttr/\tsh=0.7$, $\tin/\ttr=0.05$, $\chi=0.018$, and $\pc=15$.
(b) The same for $\ttr/\tsh=0.3$, $\tin/\ttr=2$, $\chi=0.042$ and $\pc=15$.
}
\vspace{-0.15 in}
\label{fig4}
\end{figure}

We confirm the effect of heating by analyzing the slope $\partial_\eac({ \delta \rho_\omega }/{\rho_{D} })$ of the MIRO signal at integer $\eac=N=1,2,\ldots$.
According to Eqs.~(\ref{c5a})-(\ref{gammainmodel=}), the photoresponse near integer $\eac$ should remain linear in $\pc$ at arbitrary power, as long as $\lambda$ and $\eta$ remain $\pc$-independent.
The slope is most easily found from \req{eq.miro},
\be\label{slope}
S_N \equiv \partial_\eac({ \delta \rho_\omega }/{\rho_{D} })|_{\eac=N}\simeq -4\pi^2 N \eta \pc \lambda^{2}.
\ee
Figure~\ref{fig3}(b) demonstrates that the slope $S_N$, obtained from our data at $\eac = N = 1, 2,$ and 3, indeed increases linearly with power up to $\pc\approx 1.5$ ($P/P_{\max}\approx 0.1$) but tends to saturate or even decrease at higher $P$.
Deviations at higher $\pc$ become progressively stronger for larger $N$ confirming that $\eta_\text{in}\ll \eta$ \citep{note:errors} so that heating of electrons primarily affects $\tq$ entering $\lambda^2 = \exp(-N/f\tq)$.

We now summarize our analysis procedure.
We first extract $\pc$ from the position of the satellite extrema observed at higher powers and then find other $\pc$ using known attenuation factors.
Next, by doing standard Dingle analysis on the low power data [Eq.(1)], we obtain the quantum lifetime $\tq$.
Using the inelastic rate $\tin$ from theoretical estimates, the scattering rate of smooth disorder is obtained as $1/\tsm = 1/\tq - 1/\tin$.
We then fit the low power photoresistance with Eq.(1) and (7), leaving $\tau/\tsh$ as the only fitting parameter.
The photoresistance with fixed $\tsh$ and $\tsm$ is then calculated for all power levels, with the only parameter allowed to vary with power being the electron temperature. We use the slope of photoresistance at integer $\eac$ as a crosscheck, to confirm that the inelastic contribution is small and that the primary effect of heating is the suppression of $\tau_q$.

Figure \ref{fig4}(a) illustrates the behavior of the subleading contributions $\F_\text{dis}^\text{(sm)}$ and $\F_\text{in}$ which are largely irrelevant for the conditions of our experiment but can strongly affect the photoresponse in other parametric regimes.
The contributions $\F_\text{dis}^\text{(sh)}$, $\F_\text{dis}^\text{(sm)}$, and $\F_\text{in}$ are plotted with the above parameters found for $P=P_\text{max}$, together with their sum $\F_\text{tot}=\F_\text{dis}+\F_\text{in}$ entering \req{c5a}.
Here, both $\F_\text{dis}^\text{(sm)}$ and $\F_\text{in}$ are small compared to $\F_\text{dis}^\text{(sh)}$.
The large sharp disorder component produces a pronounced fine structure in the inelastic term similar to $\F_\text{dis}^\text{(sh)}$.
Apart from being small due to $\tin/\ttr\ll 1$, $\F_\text{in}$ exhibits different positions of the secondary extrema and decays faster than $\F_\text{dis}^\text{(sh)}$ for $\xi\gg 1$.
Figure \ref{fig4}(b) illustrates the parametric regime when the inelastic time is large, $\tin/\ttr=2$, while the transport rate is dominated by the smooth component of disorder, $\chi\ttr/\tsm=0.7$ (to keep $\tqdis\simeq\tsm$ unchanged, we use $\chi=0.042$).
Here, smooth disorder washes out the fine structure in $\F_\text{in}$ which dominates the photoresponse.
Despite still significant contribution of sharp disorder to the transport rate $1/\ttr$, $\ttr/\tsh=0.3$, in this parametric regime the resulting fine structure in $\F_\text{dis}^\text{(sh)}$ is hardly visible on top of stronger, yet featureless contribution of $\F_\text{dis}^\text{(sm)}+\F_\text{in}$.

According to the above theory, a sufficiently strong short-range disorder component is crucial for the observation of the fine structure.
Apart from that, we have seen that the photoresponse is affected by detrimental heating effects which, in principle, can exponentially suppress MIRO at microwave intensities still insufficient to detect the fine structure.
To avoid excess heating, it is therefore desirable to employ lower radiation frequencies as $\pc \sim f^{-4}$.
Using lower $f$, however, inevitably pushes MIRO to lower $B$ which calls for samples with long quantum lifetimes.
In addition, high transport mobility is also desirable since it helps to reduce both the heating effects and the inelastic contribution, which otherwise could overwhelm the MIRO signal and mask the fine structure.
Finally, a nonuniformity of the microwave field across the sample may also smear out the fine structure leaving exclusively the inhomogeneously broadened primary extrema in the observed averaged signal.

In summary, we have observed and investigated high-intensity MIRO exhibiting multiple satellite oscillations.
This fine structure is qualitatively distinct from conventional MIRO and can be quantitatively explained by theory considering multiphoton-assisted scattering by short-range impurities.
Unique properties of the fine structure enable us to estimate all experimental parameters, including radiation intensity, draw quantitative conclusions about partial contributions to microwave photoresistance, and evaluate the role of electron heating.
Furthermore, fine structure offers an opportunity to quantify short- and long-range disorder contributions to the electron mobility.
For our 2DEG we demonstrate that the mobility is limited by short-range disorder, which is responsible for about 70\% of the transport scattering rate.

\begin{acknowledgments}
We thank M. Khodas for discussions.
The work at the University of Minnesota was funded by the NSF Grant No. DMR-1309578.
The work at the University of Regensburg was funded by the German Research Foundation (DFG).
The work at Princeton University was funded by the Gordon and Betty Moore Foundation through the EPiQS initiative Grant GBMF4420, and by the National Science Foundation MRSEC Grant DMR-1420541.
Preliminary measurements were performed at the National High Magnetic Field Laboratory, which is supported by NSF Cooperative Agreement No. DMR-0654118 and by the State of Florida.
\end{acknowledgments}

\clearpage


\begin{thebibliography}{49}
\bibitem[{not({\natexlab{d}})}]{note:mixed}
\bibinfo{note}{Residual impurities lead to almost isotropic scattering and are modelled as a Gaussian white-noise disorder.
The restrictive condition on the smooth component (fullfilled in our experiments) is that the distance to remote donors should be smaller than the magnetic length.}
\bibitem[{not({\natexlab{b}})}]{note:omitted}
\bibinfo{note}{In Eqs.~(\ref{eq.miro}) and (\ref{lowP})-(\ref{slope}), we use
  $\F_\text{dis}\simeq \eac\tau\partial_\eac \bar{\gamma}(\xi)$ while omitting
  the remaining part $[\tau \bar{\gamma}(\xi)-1]$ which is relatively small for
  $2\pi\eac>1$. The omitted part is included in the calculated photoresponse in
  Figs.~\ref{fig2} and \ref{fig4}, and provides a non-vanishing $\F_\text{dis}$
  at half-integer $\eac$.}
\bibitem[{not({\natexlab{b}})}]{note:errors}
\bibinfo{note}{We estimate a possible error in the extracted value of $\tau/\tau_\text{sh}\simeq 0.7$ to be $\sim 10~\%$.
Despite the accuracy of the theoretical estimate for $\tau_\text{in}$ can be lower, the associated error in $\tau/\tau_\text{sh}$ is small as long as $\eta_\text{in}\ll\eta_\text{sh}$.}
\end{thebibliography}
\end{document}